\documentclass[a4paper,draft]{amsproc}
\usepackage{amssymb}
\usepackage{amscd} 
\usepackage[dvips]{graphicx} 
\usepackage{subfigure}
\usepackage{verbatim}
\usepackage{multicol}

\theoremstyle{plain}

\theoremstyle{definition}

\theoremstyle{remark}
 
 \numberwithin{equation}{section}

\renewcommand{\leq}{\leqslant}
\renewcommand{\geq}{\geqslant}

\title{A NEW MODEL OF NONLOCAL MODIFIED GRAVITY}



\author{Ivan Dimitrijevic}

\address{
Faculty of Mathematics, 
University  of Belgrade, 
Belgrade,
Serbia} 

\author{Branko Dragovich}

\address{
Institute of Physics, 
University  of Belgrade, 
Belgrade,
Serbia} 

\author{Jelena Grujic}

\address{
Teachers Training Faculty, 
University  of Belgrade, 
Belgrade,
Serbia} 

\author{ Zoran Rakic}

\address{
Faculty of Mathematics, 
University  of Belgrade, 
Belgrade,
Serbia}

\thanks{\,Work partially supported by the Serbian Ministry of Education, Science and Technological Development, contract No. 174012} 

\newcommand{\Real}{\mathbb{R}}

\newcommand{\pmu}{\partial_{\mu}}
\newcommand{\pnu}{\partial_{\nu}}
\newcommand{\nmu}{\nabla_{\mu}}
\newcommand{\nnu}{\nabla_{\nu}}

\newcommand{\FF}{\mathcal{F}}

\begin{document}


\begin{abstract}
We consider a new modified gravity model with nonlocal term of the form $R^{-1} \mathcal{F}(\Box) R. $
This kind of nonlocality is motivated by investigation of applicability of a few unusual ans\"atze to obtain
some exact cosmological solutions. In particular, we find attractive and useful quadratic ansatz $\Box R = q R^{2}.$
\end{abstract}

\maketitle

\section{Introduction}
\label{sec:1}

 In spite of the great successes of General  Relativity (GR) it has not got status of a complete theory of gravity. To modify GR there are motivations coming from its quantum aspects, string theory,  astrophysics and cosmology. For example, cosmological solutions of GR contain Big Bang singularity, and Dark
Energy as a cause for accelerated expansion of the Universe. This initial cosmological singularity is an evident signature that GR is not appropriate theory of the Universe at cosmic time $t=0.$ Also, GR has not been verified at the very large cosmic scale and dark energy has not been discovered in the laboratory experiments. This situation gives rise to research for an adequate modification of GR among numerous possibilities (for a recent review, see \cite{clifton}).

Recently it has been shown that nonlocal modified gravity with action
\begin{equation} \label{eq-1.1}
S =  \int d^{4}x \sqrt{-g}\Big(\frac{R - 2 \Lambda}{16 \pi G} + C R \mathcal{F}(\Box) R  \Big),
\end{equation}
 where $R$ is scalar curvature, $\Lambda$ -- cosmological constant, $ \mathcal{F}(\Box)= \displaystyle \sum_{n =0}^{\infty} f_{n}\Box^{n}$  is an analytic function  of the d'Alembert-Beltrami operator $\Box = \frac{1}{\sqrt{-g}} \partial_{\mu}
\sqrt{-g} g^{\mu\nu} \partial_{\nu},$ $\, g = det(g_{\mu\nu}) $ and $C$ is a
constant, has nonsingular bounce cosmological solutions, see \cite{biswas0,biswas,koshelev,dragovich}. To solve equations of motion it was used ansatz
$\Box R = r R + s.$ In \cite{dragovich1} we introduced some new ans\"atze, which gave trivial solutions for the above nonlocal model \eqref{eq-1.1}.
In this paper we consider some modification of the above action in the nonlocal sector, i.e.
\begin{equation} \label{eq-1.2}
S =  \int d^{4}x \sqrt{-g}\Big(\frac{R}{16 \pi G} +  R^{-1} \mathcal{F}(\Box) R  \Big)
\end{equation}
and look for nontrivial cosmological solutions for the new ans\"atze  (see \cite{dragovich1}). Note that the cosmological constant $\Lambda$ in \eqref{eq-1.2}
is hidden in the term  $f_0,$ i.e. $\Lambda = - 8 \pi G f_0.$  To the best of our knowledge action (\ref{eq-1.2}) has not been considered so far. However, there are investigations of gravity modified by $1/R$ term  (see, e.g. \cite{Woodard} and references therein), but it is without nonlocality.\\

\section{Equations of motion}
\label{sec:2}

 By variation of
action \eqref{eq-1.2} with respect to metric $g^{\mu\nu}$ one
obtains the equations of motion for $g_{\mu\nu}$
\begin{equation} \begin{aligned}\label{eq-2.1}
&R_{\mu\nu} V - (\nmu\nnu - g_{\mu\nu} \Box)V -
\frac{1}{2} g_{\mu\nu} R^{-1} \mathcal{F}(\Box) R \\
&+ \sum_{n=1}^{\infty} \frac{f_n}{2} \sum_{l=0}^{n-1} \big(
g_{\mu\nu} \left( \partial_{\alpha} \Box^l(R^{-1})
\partial^{\alpha} \Box^{n-1-l} R + \Box^l(R^{-1}) \Box^{n-l} R
\right)  \\
&- 2 \pmu \Box^l(R^{-1}) \pnu \Box^{n-1-l} R\big)   = - \frac{G_{\mu\nu}}{16 \pi G} ,\\
&V = \FF(\Box) R^{-1} - R^{-2} \FF(\Box) R .
\end{aligned} \end{equation}

The trace of the equation \eqref{eq-2.1} is
\begin{equation}\begin{aligned}\label{eq-2.2}
&R V + 3 \Box V + \sum_{n=1}^{\infty} f_n \sum_{l=0}^{n-1} \left( \partial_{\alpha} \Box^l(R^{-1}) \partial^{\alpha} \Box^{n-1-l} R + 2 \Box^l(R^{-1}) \Box^{n-l} R \right)  \\
&-2 R^{-1} \FF(\Box) R  = \frac{R}{16 \pi G}.
\end{aligned} \end{equation}

The $00$ component of \eqref{eq-2.1} is
\begin{equation} \begin{aligned}\label{eq-2.3}
&R_{00} V - (\nabla_0\nabla_0 - g_{00} \Box)V -
\frac{1}{2} g_{00} R^{-1} \mathcal{F}(\Box) R \\
&+ \sum_{n=1}^{\infty} \frac{f_n}{2} \sum_{l=0}^{n-1} \big(
g_{00} \left( \partial_{\alpha} \Box^l(R^{-1})
\partial^{\alpha} \Box^{n-1-l} R + \Box^l(R^{-1}) \Box^{n-l} R \right)  \\
&- 2 \partial_0 \Box^l(R^{-1}) \partial_0 \Box^{n-1-l} R\big)  = - \frac{G_{00}}{16 \pi G}.
\end{aligned} \end{equation}

We use Friedmann-Lema\^{\i}tre-Robertson-Walker (FLRW) metric
$ds^2 = - dt^2 + a^2(t)\big(\frac{dr^2}{1-k r^2} + r^2 d\theta^2 +
r^2 \sin^2 \theta d\phi^2\big)$ and investigate all three
possibilities for curvature parameter $k$ ($0,\pm 1$). In the FLRW
metric scalar curvature is $R = 6 \left (\frac{\ddot{a}}{a} +
\frac{\dot{a}^{2}}{a^{2}} + \frac{k}{a^{2}}\right )$ and $\Box h(t)=
- \partial_t^2 h(t) - 3 H \partial_t h(t) ,$ where $H =
\frac{\dot{a}}{a}$ is the Hubble parameter.   In the sequel we
shall use three kinds of ans\"atze (two of them introduced in
\cite{dragovich1}) and solve equations of motions \eqref{eq-2.2}
and \eqref{eq-2.3} for cosmological scale factor in the form $a(t)
= a_0 |t-t_0|^\alpha .$
\\

\section{Quadratic ansatz: $\Box R = q R^2$}

Looking for solutions in the form $a(t) = a_0 |t-t_0|^{\alpha}$
this ansatz becomes

\begin{equation}
\begin{aligned} \label{3.1}
&\alpha (2\alpha - 1)(q \alpha(2\alpha - 1) - (\alpha - 1)) (t-t_0)^{-4} \\
&+ \frac{\alpha k}{3a_0^2} (1-\alpha + 6q(2\alpha - 1))
(t-t_0)^{-2\alpha - 2} + \frac{q k^2}{a_0^4}(t-t_0)^{-4\alpha} = 0.
\end{aligned}
\end{equation}

Equation \eqref{3.1} is satisfied for all values of time $t$ in six cases:

\begin{multicols}{2}
\begin{enumerate}
\item $k=0$, $\alpha = 0$, $q\in \Real$,
 \item $k=0$, $\alpha=\frac 12$, $q\in \Real$,
 \item $k=0$, $\alpha \neq 0$ and $\alpha \neq \frac 12$, $q = \frac{\alpha-1}{\alpha(2\alpha -1)}$,
\item $k=-1$, $\alpha = 1$, $q\neq 0$, $a_0=1$,
 \item $k\neq 0$, $\alpha = 0$, $q=0$,
 \item $k\neq 0$, $\alpha = 1$, $q=0$.
\end{enumerate}
\end{multicols}

In the cases (1), (2) and (4) we have $ R=0$ and therefore $R^{-1}$ is not defined. The case (5) yields a solution
which does not satisfy equations of motion. Hence there remain two cases for further consideration.    \

\subsection{Case $k=0, \, q = \frac{\alpha-1}{\alpha(2\alpha -1)}$}

For this case, we have the following expressions  depending on the  parameter $\alpha:$
\begin{equation}
\begin{aligned}
q &= \frac{\alpha-1}{\alpha(2\alpha -1)}, \, &R &= 6\alpha(2\alpha-1)(t-t_0)^{-2},\\
a &= a_0 |t-t_0|^{\alpha}, \,  & H &= \alpha (t-t_0)^{-1} , \\
R_{00}& = 3\alpha (1-\alpha) (t-t_0)^{-2}, \, & G_{00}&=
3\alpha^{2}(t-t_{0})^{-2}.
\end{aligned}
\end{equation}

We now express $\Box^n R$ and $\Box^n R^{-1}$
in the following way:
\begin{equation}
\begin{aligned}
\Box^n R &= B(n,1) (t-t_0)^{-2n-2} , \quad
\Box^n R^{-1} = B(n,-1) (t-t_0)^{2-2n} ,\\
B(n,1) &= 6\alpha(2\alpha-1) (-2)^{n} n!\prod_{l=1}^{n}(1-3\alpha +2l) ,  \; n \geq 1 ,\\
B(n,-1) &= (6\alpha(2\alpha-1))^{-1} 2^{n}\prod_{l=1}^{n}(2-l)(-3-3\alpha +2l) , \; n \geq 1 ,\\
B(0,1) &= 6\alpha(2\alpha-1), \quad
B(0,-1) = B(0,1)^{-1} .   
\end{aligned}
\end{equation}

Note that  $ B(1,-1) = - \frac{3\alpha +1}{3\alpha(2\alpha-1)} =
-2 (3\alpha +1) B(0,1)^{-1} $ and $B(n,-1) =0$ if $n \geq 2 .$ Also,
we obtain
\begin{equation}
\begin{aligned}
\mathcal{F}(\Box) R &= \sum_{n=0}^{\infty} f_n B(n,1)
(t-t_0)^{-2n-2},\\
\mathcal{F}(\Box) R^{-1} &= f_0 B(0,-1)(t-t_0)^2 + f_1 B(1,-1) .
\end{aligned}
\end{equation}

Substituting these equations into trace  and 00 component of the EOM one has
\begin{equation} \label{eom:qr2k0}
\begin{aligned}
&r^{-1} \sum_{n=0}^{\infty} f_n B(n,1) \left( -3 r  + 6 (1-n)(1-2n+3\alpha)\right) (t-t_0)^{-2n} \\
&+ r\sum_{n=0}^{1} f_n \left(r B(n,-1) + 3 B(n+1,-1) \right) (t-t_0)^{-2n} \\
&+ 2 r \sum_{n=1}^{\infty} f_n \gamma_n (t-t_0)^{-2n}  =
\frac{r^2}{16\pi G} (t-t_0)^{-2} , \\
& \sum_{n=0}^{\infty} f_n r^{-1} B(n,1) \left(\frac{r}{2} - A_n \right) (t-t_0)^{-2n} \\
&+ \sum_{n=0}^{1} f_n r B(n,-1)\, A_n  \, (t-t_0)^{-2n} +
\frac{r}{2}  \sum_{n=1}^{\infty} f_n \delta_n (t-t_0)^{-2n}
\\  &= \frac{-r^2}{32\pi G}\frac{\alpha}{2\alpha-1}(t-t_0)^{-2},
\end{aligned}
\end{equation}
where $r = B(0,1)$ and
\begin{equation}
\begin{aligned}
\gamma_n &= \sum_{l=0}^{n-1} B(l,-1)(B(n-l,1) + 2(1-l)(n-l) B(n-l-1,1)), \\
\delta_n &= \sum_{l=0}^{n-1} B(l,-1)(- B(n-l,1) + 4(1-l)(n-l) B(n-l-1,1)), \\
A_n &= 6\alpha(1-n) - r \frac{\alpha - 1}{2(2\alpha - 1)} =
\frac{r}{2} \, \frac{3-2n -\alpha}{2\alpha -1}.
\end{aligned}
\end{equation}

Equations  \eqref{eom:qr2k0} can be split into system of pairs of equations with respect to each coefficient $f_n$. In the case $n>1,$ there are the following pairs:
\begin{equation} \label{sys1}
\begin{aligned}
 f_n& \big(  B(n,1) \left( -3 r  + 6 (1-n)(1-2n+ 3\alpha)\right) + 2 r^2 \gamma_n \big)  = 0 , \\
  f_n& \Big( B(n,1) \left( \frac{r}{2} - A_n \right) + \frac{r^2}{2}  \delta_n \Big) = 0.
\end{aligned}
\end{equation}

Taking $\frac{3\alpha-1}{2}$ to be a natural number one obtains:
\begin{align}
B(n,1)&=6\alpha (2\alpha-1)4^{n}n!\frac{(\frac 32(\alpha-1))!}{(\frac 32(\alpha-1)-n)!} , \quad n < \frac{3\alpha-1}2,\\
B(n,1)&=0, \quad n\geq \frac{3\alpha-1}2, \\
\gamma_{n}&= 2 B(0,-1)B(n-1,1)(3n \alpha - 2 n^2 -3\alpha -1), \quad n\leq \frac{3\alpha-1}2, \\
\delta_{n}&= 2B(0,-1)B(n-1,1)(2n^2 + 3n+ 3\alpha -3\alpha n+1), \quad n\leq \frac{3\alpha-1}2, \\
\gamma_n &= \delta_{n}=0, \quad n>\frac{3\alpha-1}2 .
\end{align}

If $n>\frac{3\alpha-1}2$ then $B(n,1) = \gamma_n = \delta_n = 0$ and hence the system is trivially satisfied for arbitrary value
of coefficients $f_n$. On the other hand for $ 2 \leq n \leq \frac{3\alpha-1}2$ the system has only trivial solution $f_n=0$.

When $n=0$ the pair becomes
\begin{align} \label{sys2}
f_0 \big(  -2 r + 6(1+3\alpha) + 3 r B(1,-1) \big)  = 0, \quad f_0  = 0
\end{align}
and its solution is $f_0 = 0$. The remaining case $n=1$ reads
\begin{equation} \label{sys3}
\begin{aligned}
&f_1 \big(- 3r^{-1}B(1,1) + r B(1,-1)  + 2 \gamma_1 \big)  = \frac{r}{16\pi G}, \\
& f_1 \Big( A_1 (r B(1,-1) - r^{-1}B(1,1)) + \frac{1}{2}(B(1,1)+ r
\delta_1) \Big) = \frac{-r^2}{32\pi G} \; \frac{\alpha}{2\alpha -
1},
\end{aligned}
\end{equation}
and it gives $f_1 = -\frac{3\alpha(2\alpha-1)}{32 \pi G (3\alpha
-2)}$. \

\subsection{Case $k\neq 0$, $\alpha = 1$, $q=0$}

In this case
\begin{equation}\begin{aligned}
a &= a_0 |t-t_0|, \quad H = (t-t_0)^{-1},  \quad R = s(t-t_0)^{-2}, \\
s &=6(1+ \frac{k}{a_0^2}), \quad \Box R = 0, \quad R_{00}  = 0, \\
\Box^n R^{-1} &= D(n,-1) (t-t_0)^{2-2n}, \\
D(0,-1) &= s^{-1}, \quad  D(1,-1) = -8 s^{-1}, \quad D(n,-1) =0, \; n\geq 2.
\end{aligned} \end{equation}

Substitution of the above expressions in trace and 00 component of the EOM yields
\begin{equation}
\begin{aligned} \label{30.04:1}
& 3 f_0  + \sum_{n=0}^1 f_n s D(n,-1) (t-t_0)^{-2n} + 4 f_1  (t-t_0)^{-2} = \frac{s}{16 \pi G} (t-t_0)^{-2},  \\
& -6 f_0 s^{-1} + \frac{1}{2} f_0 + 6 \sum_{n=0}^{1} f_n D(n,-1) (1-n) (t-t_0)^{-2n} \\
&+ 2 f_1  (t-t_0)^{-2} = - \frac{s}{32 \pi G} (t-t_0)^{-2}.
\end{aligned}
\end{equation}

This system leads to  conditions for $f_0$ and $f_1 :$
\begin{equation} \begin{aligned}
-2 f_0 - 4 f_1 (t-t_0)^{-2} &= \frac{s}{16 \pi G} (t-t_0)^{-2}, \\
\frac{1}{2} f_0 + 2 f_1 (t-t_0)^{-2} &=  - \frac{s}{32 \pi G} (t-t_0)^{-2}.
\end{aligned} \end{equation}

The corresponding solution  is
\begin{equation} 
f_0 = 0, \quad  f_1 = \frac{-s}{64 \pi G}, \quad f_n \in \Real, \quad n\geq 2.
\end{equation}
\\

\section{Ansatz $\Box^{n}R = c_{n}R^{n+1}, \, \, n \geq 1$}
Presenting
$\Box^{n+1} R$ in two ways:
\begin{equation} \begin{aligned}
&\Box^{n+1} R = \Box c_n R^{ n + 1} \nonumber \\
& = c_n ((n + 1) R^{n} \Box R - n( n + 1)R^{ n - 1} \dot{R}^2) \nonumber \\
& = c_n (n + 1) (c_1 R^{ n + 2} - n R^{n  - 1} \dot{R}^2) \nonumber  \\
& = c_{n+1} R^{ n + 2}   \label{23.03:1}
\end{aligned} \end{equation}
 it follows
\begin{align}
&\dot{R}^2 = R^{3},  \label{23.03:2} \\
&c_{n+1} = c_n (n + 1)(c_1 -  n ), \label{23.03:3}
\end{align}
where $\dot{R}^2$ means $(\dot R)^2$.

General solution of equation  \eqref{23.03:2} is
\begin{equation}
R = \frac{4}{(t-t_0)^{2}},\; \; t_{0} \in \mathbb{R}.
\label{23.03:4}
\end{equation}
Taking $ n = 1 $ in the  ansatz  yields
\begin{equation}
\Box R = c_{1}R^{2}. \label{23.03:5}
\end{equation}
Substitution of \eqref{23.03:4} in \eqref{23.03:5} gives $H = \frac{2c_{1}+3}{3(t-t_0)}$. This implies
\begin{equation}
a(t) = a_{0} |t- t_0|^{\frac{2c_{1}+3}{3}}, \quad  a_{0}> 0. \label{23.03:6}
\end{equation}
Using \eqref{23.03:4} in equation
\begin{equation}
R = 6 \left (\frac{\ddot{a}}{a} + \frac{\dot{a}^{2}}{a^{2}} +
\frac{k}{a^{2}}\right ) \label{23.03:7}
\end{equation}
gives
\begin{equation}
(t-t_0)^{2}\ddot{y}- \frac{4}{3}y = -2k(t-t_0)^{2}, \;
\mbox{where }  y = a^{2}(t). \label{23.03:8}
\end{equation}
It  can be shown that general solution of the last equation is
\begin{equation}
a^{2}(t)= \tilde{C_{1}}|t-d_{1}|^{\frac{3+\sqrt{57}}{6}} +
\tilde{C_{2}}|t-d_{1}|^{\frac{3-\sqrt{57}}{6}}-3k|t-d_{1}|^{2}, \;
\; \tilde{C_{1}}, \; \tilde{C_{2}} \in \mathbb{R}. \label{23.03:9}
\end{equation}
By comparison of the last equation with \eqref{23.03:6} one can conclude:
\begin{enumerate}
\item  If $ c_{1}=0 $ then $ k $ must be equal to $ -1 $. In this
case $ \Box^{n}R=0, \quad  n \geq 1 $.

\item  If $ c_{1}\neq 0$ then $ k $ must be equal to $ 0 $. In
this case $ c_{1} = \frac{-9 \pm \sqrt{57}}{8} $.
\end{enumerate}
\

\subsection{Case $\Box^{n}R = c_{n}R^{n+1}, \,  c_{1}=0$}
From the previous analysis, it follows:

\begin{equation} \label{4.04:4}
\begin{aligned}
k &=-1, \quad
& a(t) &= \sqrt{3}|t-t_0|, \quad
&H(t)&= \frac{1}{t-t_0},\\
R &= \frac{4}{(t-t_0)^{2}}, \quad
&\Box^{n}R &= 0, \quad n \geq 1, \quad
& \mathcal{F}(\Box) R &= f_{0}R.
\end{aligned}
\end{equation}

It can be shown that
\begin{align} \label{4.10}
\Box^{n}R^{-1}= (-1)^n 4^{n-1} \prod_{l=0}^{n-1} (1-l)(2-l)(t-t_0)^{2-2n}.
\end{align}

From \eqref{4.10} follows $ \Box^{n}R^{-1}= 0, \, n > 1 $. Then
\begin{align}\label{4.04:5}
\mathcal{F}(\Box) (R^{-1})= f_{0}R^{-1} + f_{1}\Box R^{-1}.
\end{align}

Substituting  \eqref{4.04:4} and \eqref{4.04:5} in the $ 00 $
component of the EOM one obtains
\begin{align}
\frac{f_{0}}{2}(t-t_0)^{2} + 2f_{1} + \frac{1}{8 \pi G}= 0
\end{align}
and it follows
\begin{align} \label{4.13}
f_{0} = 0, \quad f_{1} = \frac{-1}{16 \pi G}, \quad
f_n \in \Real, \quad n \geq 2.
\end{align}

Substituting  \eqref{4.04:4} and  \eqref{4.04:5} in the trace
equation one has
\begin{align}
-2f_{0} (t-t_0)^{2} - 4f_{1} -\frac{1}{4 \pi G}= 0
\end{align}
and it gives the same  result \eqref{4.13}.

\subsection{Case $\Box^{n}R = c_{n}R^{n+1}, \, c_{1} = \frac{-9 \pm \sqrt{57}}{8}$}

In this case:
\begin{align}
k &= 0, \quad R = \frac{4}{(t-t_0)^{2}}, \quad
H = \frac{2c_1+3}{3(t-t_0)}, \quad
a = a_{0}|t-t_{0}|^{\frac{2c_1+3}{3}},\; \; a_{0}> 0,\\
R_{00} &= 3 \alpha(1-\alpha)(t-t_0)^{-2}, \, \,
G_{00} = (3 \alpha(1-\alpha)+ 2)(t-t_0)^{-2}, \, \, \alpha = \frac{2c_{1}+3}{3}, \nonumber \\
\Box^{n}R &=  4^{n+1}c_{n}(t-t_0)^{-2n-2}, \quad c_{0}=1 . \nonumber
\end{align}

One can show that
\begin{align}
\Box^{n}R^{-1} &= M(n,-1)(t-t_0)^{2-2n},
\end{align}
where
\begin{align}
M(0,-1) &= \frac{1}{4}, \quad
&M(1,-1) &=-(c_1+2), \quad
&M(n,-1) &= 0, \quad n > 1.
\end{align}
Also one obtains
\begin{equation}
\begin{aligned} \label{4.18}
\mathcal{F}(\Box) R &= \sum_{n=0}^{\infty}4^{n+1} f_n c_n (t-t_0)^{-2n-2}, \\
\mathcal{F}(\Box) R^{-1} &= f_{0}M(0,-1)(t-t_0)^{2}+ f_{1}M(1,-1).
\end{aligned}
\end{equation}

Substituting \eqref{4.18} in the trace equation it becomes
\begin{align} \label{4.19}
&- \frac{1}{4\pi G} (t-t_0)^{-2} -2f_{0}- 3\sum_{n=1}^{\infty}4^{n} f_n c_n (t-t_0)^{-2n} \nonumber \\
&+  \sum_{n=1}^{\infty} f_n  \big(\sum_{l=0}^{n-1}
M(l,-1)4^{n-l+1}((1-l)(n-l)c_{n-1-l}+ 2c_{n-l})\big)
(t-t_0)^{-2n} \\
&+f_1(4M(1,-1)+ 3M(2,-1))(t-t_0)^{-2} = 0. \nonumber
\end{align}

To satisfy  equation \eqref{4.19} for all values of time $t$ one obtains:
\begin{align}
&f_{0} = 0, \quad f_{1}(2c_{1}+1) = -\frac{1}{16 \pi G},
\end{align}
\begin{align}
&f_{n}\big(-3 c_{n}+\sum_{l=0}^{1} M(l,-1)4^{1-l}((1-l)(n-l)c_{n-1-l}+ 2c_{n-l})\big) =0, \; n \geq 2.
\end{align}
Suppose that $f_n \neq 0$ for $n \geq 2$, then from the last equation follows
\begin{align}
-3 c_{n}+\sum_{l=0}^{1} M(l,-1)4^{1-l}((1-l)(n-l)c_{n-1-l}+ 2c_{n-l})=0
\end{align}
and it becomes
\begin{align} \label{4.23}
c_{n-1}(n^{2}-c_{1}n-2c_{1}-4)=0.
\end{align}
Since $ c_{n-1} \neq 0,$ condition \eqref{4.23} is satisfied for $ n=-2$
or $ n= c_{1}+2$.\\
Hence, we  conclude that $f_n = 0$ for $n \geq 2$.\\ \\

Since $f_n=0$ for $n \geq 2$, the 00 component of the EOM becomes
\begin{equation}
\begin{aligned} \label{4.24}
&\frac{1}{16 \pi G}(-3 \alpha^{2}+ 3 \alpha +2)(t-t_0)^{-2} \\
& + \frac{1}{2}f_{0}(\frac{3}{2}\alpha^{2}- \frac{9}{2}\alpha + 1)+ f_{1}c_{1}(3 \alpha^{2}- 3 \alpha +2)(t-t_0)^{-2}\\
&+ 8 f_{1} M(0,-1)(1-c_{1})(t-t_0)^{-2}+ 3\alpha(3-\alpha)M(0,-1)f_{0} \\
&-3\alpha(\alpha-1)M(1,-1)f_{1}(t-t_0)^{-2}=0.
\end{aligned}
\end{equation}

In order to satisfy  equation \eqref{4.24} for all values of time $t$ it has to be
\begin{align}
f_{0}=0, \quad f_{1}(\frac{4}{3}c_{1}^{3}+ \frac{10}{3}c_{1}^{2}+
2c_{1} +1) = \frac{1}{16 \pi G }(\frac{2}{3}c_{1}^{2} + c_{1} -1).
\end{align}

The necessary and sufficient condition for the EOM to have a solution is
\begin{align}
c_{1} (8 c_1^2 + 18 c_1 + 3 )= 0.
\end{align}

Since  $c_{1} = \frac{-9\pm\sqrt{57}}{8},$ the last condition is satisfied.
\\

\section {Cubic ansatz: $\Box R = q R^3$}
Recall that we are looking for solutions in the form $a(t) = a_0 |t-t_0|^\alpha$. In the explicit form it reads
\begin{equation} \begin{aligned} \label{qubic:1}
&\alpha(\alpha  - 1) \Big( 3(2\alpha -1) (t-t_0)^{-4} + \frac {k}{a_0^2}  (t-t_0)^{-2\alpha -2}  \Big) \\
&= 18 q \Big( \alpha(2\alpha-1) (t-t_0)^{-2} + \frac k{a_0^2}
(t-t_0)^{-2\alpha}\Big)^3.
\end{aligned} \end{equation}

It yields the following seven possibilities:
\begin{multicols}{2}
\begin{enumerate}
\item $k=0$, $\alpha = 0$, $q\in \Real$, \label{cub:1} \item
$k=0$, $\alpha = \frac 12$, $q\in \Real$,\label{cub:2} \item
$k=-1$, $\alpha = 1$, $q\neq 0$, $a_0=1$,\label{cub:3} \item
$k=0$, $\alpha = 1$, $q=0$,\label{cub:4} \item $k\neq 0$, $\alpha
= 0$, $q=0$,\label{cub:5} \item $k\neq 0$, $\alpha = 1$,
$q=0$,\label{cub:6} \item $k\neq 0$, $\alpha = \frac 12$,
$q=-\frac{a_0^4}{72}$. \label{cub:7}
\end{enumerate}
\end{multicols}

Cases \eqref{cub:1}, \eqref{cub:2} and \eqref{cub:3} contain scalar curvature $R = 0,$ and therefore we will not  discuss them. Cases \eqref{cub:4}, \eqref{cub:5} and \ref{cub:6} are also obtained from the quadratic ansatz and have been discussed earlier.
The last case contains:
\begin{equation} \begin{aligned}
a(t) &= a_0 \sqrt{|t-t_0|}, \quad
H(t) = \frac{1}{2(t-t_0)}, \\
R(t) &= \frac{6k}{a_0^{2}}|t -t_0|^{-1}, \quad R_{00} =
\frac{3}{4(t-t_0)^{2}}.
\end{aligned} \end{equation}

One can derive the following expressions:
\begin{equation}
\begin{aligned}  \label{5.3}
\Box^{n}R &= N(n,1)|t-t_0|^{-2n-1},\quad
\Box^{n}R^{-1}  = N(n,-1) |t-t_0|^{1-2n}, \\
N(0,1) &=\frac{6k}{a_0^{2}},\quad  N(0,-1) = N(0,1)^{-1}, \\
N(n,1) &= N(0,1) (-1)^n \prod_{l=0}^{n-1} (2l+1)(2l+\frac 12), \quad n\geq 1, \\
N(n,-1) &= N(0,1)^{-1} (-1)^n \prod_{l=0}^{n-1} (2l-1)(2l-\frac 32), \quad n\geq 1, \\
\mathcal{F}(\Box) R &= \sum_{n=0}^{\infty} f_n
N(n,1)|t-t_0|^{-2n-1}, \\
\mathcal{F}(\Box) R^{-1} &= \sum_{n=0}^{\infty} f_n
N(n,-1)|t-t_0|^{1-2n}.
\end{aligned}
\end{equation}

Substituting \eqref{5.3} in the trace equation  we obtain
\begin{equation}\begin{aligned}  \label{26.01:tr}
& -2 N(0,1)^{-1} \sum_{n=0}^\infty f_n N(n,1)|t-t_0|^{-2n} + N(0,1) \sum_{n=0}^\infty f_n (N(n,-1) \\ &- N(0,1)^{-2}N(n,1))|t-t_0|^{-2n} \\
& + 3 \sum_{n=0}^\infty f_n (N(n,-1) - N(0,1)^{-2}N(n,1))|t-t_0|^{-1-2n} \\
& + \sum_{n=1}^\infty f_n \sum_{l=0}^{n-1} N(l,-1)((1-2l)(-2n+2l+1)N(n-l-1,1) \\
& +2N(n-l,1)) |t-t_0|^{-2n}
 = \frac{N(0,1)}{16\pi G} |t-t_0|^{-1}.
\end{aligned} \end{equation}

This equation implies the following conditions on coefficient $f_0 :$
\begin{align}
f_{0} = 0, \quad \frac{N(0,1)}{16 \pi G}= 0.
\end{align}
Since $N(0,1) \neq 0$, the last equation never holds and therefore there is no solution in this case.
\\

\section{Concluding remarks}

Using a few new ans\"atze we have shown that equations of motion for nonlocal gravity model given by
action \eqref{eq-1.2} yield some bounce cosmological solutions of the form $a(t) = a_0 |t-t_0|^\alpha$.
These solutions lead to $f_0 = 0$ and hence $\Lambda = 0,$ and when $ t \rightarrow \infty $ then $ R \rightarrow 0.$
In particular, quadratic ansatz $\Box R = q R^2$ is very promising. Note that ansatz $\Box^{n}R = c_{n}R^{n+1}, \, \, n \geq 1,$
can be viewed as a special case  of ansatz $\Box R = q R^2.$

It is worth noting that equations of motion \eqref{eq-2.2} and \eqref{eq-2.3} have the de Sitter solutions $a(t) = a_0 \cosh (\lambda t), \, \,  k= +1$
and $a(t) = a_0 e^{\lambda t}, \, \, k=0,$ when $f_0 = \frac{-3\lambda^{2}}{8\pi G }  = \frac{-\Lambda}{8\pi G }, \, \,
f_n \in \Real, \, n\geq 1.$

This investigation can be generalized to some cases with $R^{-p} \mathcal{F}(\Box) R^{q}$ nonlocal term, where
$p$ and $q$ are some natural numbers satisfying $q - p \geq 0 .$ It will be presented elsewhere with discussion of various properties.

\bibliographystyle{amsplain}


\end{document}